\documentclass[letter]{aa} 
\usepackage{longtable,booktabs}
\usepackage{latexsym}
\usepackage{amssymb}
\usepackage{lscape}
\usepackage{natbib}
\usepackage{subfigure}
\usepackage{xcolor}
\usepackage{multicol}
\usepackage{multirow}
\usepackage{amsmath}
\usepackage{booktabs}
\usepackage{caption}
\usepackage{pdflscape}
\usepackage{graphicx}
\usepackage{txfonts}
\def\lsim{\mathrel{\rlap{\lower 3pt \hbox{$\sim$}} \raise 2.0pt \hbox{$<$}}}
\def\gsim{\mathrel{\rlap{\lower 3pt \hbox{$\sim$}} \raise 2.0pt \hbox{$>$}}}

\begin{document}

\title{The first blazar observed at z>6}
\subtitle{}

\author{S.Belladitta\inst{1,2}                                
\and A. Moretti\inst{1}                                                                  
\and A. Caccianiga\inst{1}
\and C. Spingola\inst{3,4} 
\and P. Severgnini\inst{1} 
\and R. Della Ceca\inst{1} 
\and G. Ghisellini\inst{5} 
\and D. Dallacasa\inst{3,4}
\and T. Sbarrato\inst{5,6}
\and C. Cicone\inst{7}
\and L. P. Cassar\`a\inst{8}
\and M. Pedani\inst{9}
}

\institute{INAF $-$ Osservatorio Astronomico di Brera, via Brera, 28, 20121 Milano, Italy\\
\email {silvia.belladitta@inaf.it}
\and
DiSAT, Universit\`a degli Studi dell'Insubria, Via Valleggio 11, 22100 Como, Italy
\and
INAF $-$ Istituto di Radioastronomia, Via Gobetti 101, I$-$40129, Bologna, Italy 
\and
Dipartimento di Fisica e Astronomia, Universit\`a degli Studi di Bologna, Via Gobetti 93/2, I$-$40129 Bologna, Italy 
\and 
INAF $-$ Osservatorio Astronomico di Brera, sede di Merate, via E. Bianchi, 46, 23807 Merate, Italy
\and 
Dipartimento di Fisica G. Occhialini, Universit\`a degli Studi di Milano Bicocca, Piazza della Scienza 3, 20126 Milano, Italy
\and
Institute of Theoretical Astrophysics, University of Oslo, P.O. Box 1029, Blindern, 0315 Oslo, Norway
\and
INAF $-$ Istituto di Astrofisica Spaziale e Fisica Cosmica (IASF), Via A. Corti 12, 20133 Milano
\and
INAF $-$ Fundaci\'on Galileo Galilei, Rambla Jos\'e Ana Fernandez P\'erez 7, 38712 Bre\~{n}a Baja, TF, Spain
}


\date{Received; accepted}

\abstract
{We present the discovery of PSO J030947.49+271757.31, the radio brightest (23.7 mJy at 1.4 GHz) active galactic nucleus (AGN) at z>6.0.
It was selected by cross-matching the NRAO VLA Sky Survey and the Panoramic Survey Telescope and Rapid Response System PS1 databases and its high-z nature was confirmed by a dedicated spectroscopic observation at the Large Binocular Telescope.
A pointed Neil Gehrels $Swift$ Observatory XRT observation allowed us to measure a flux of $\sim$3.4$\times$10$^{-14}$ erg s$^{-1}$ cm$^{-2}$ in the [0.5-10] keV energy band, which also makes this object the X-ray brightest AGN ever observed at z>6.0.      
Its flat radio spectrum ($\alpha_{\nu r}$<0.5), very high radio loudness (R>10$^3$), and strong X-ray emission, compared to the optical, support the hypothesis of the blazar nature of this source. 
Assuming that this is the only blazar at this redshift in the surveyed area of sky, we derive a space density of blazars at z$\sim$6 and with M$_{\rm 1450 \AA}$ < -25.1 of 5.5$^{+11.2}_{-4.6}$$\times$10$^{-3}$ Gpc$^{-3}$.
From this number, and assuming a reasonable value of the bulk velocity of the jet ($\Gamma$=10), we can also infer a space density of the entire radio-loud AGN population at z$\sim$6 with the same optical/UV absolute magnitude of 1.10$^{+2.53}_{-0.91}$ Gpc$^{-3}$.
Larger samples of blazars will be necessary to better constrain these estimates.
}

\maketitle

\section{Introduction}
\label{intro}
\begin{figure*}[!h]
        \centering
        {\includegraphics[width=2.2cm]{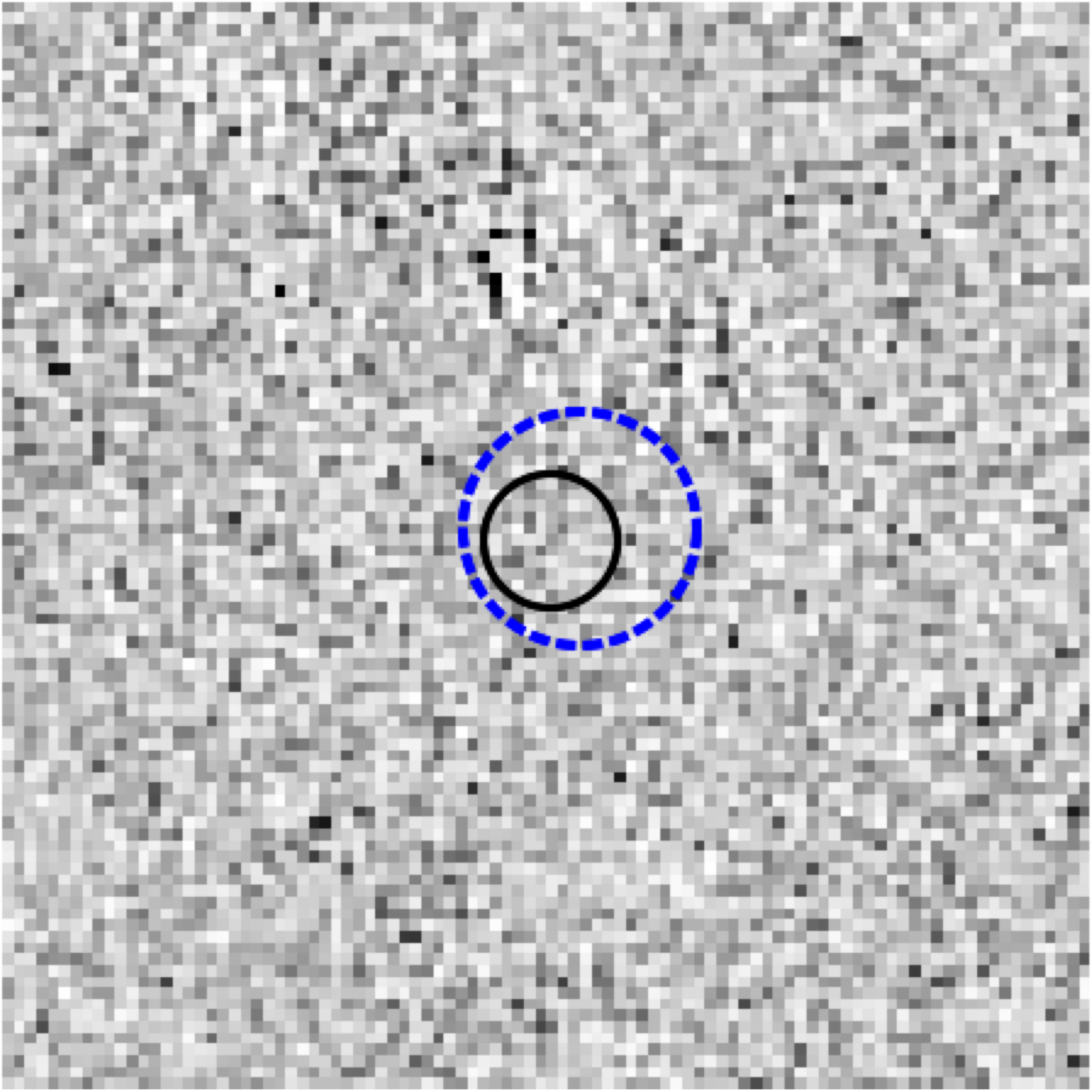}\hspace{-0.1mm}
                \includegraphics[width=2.2cm]{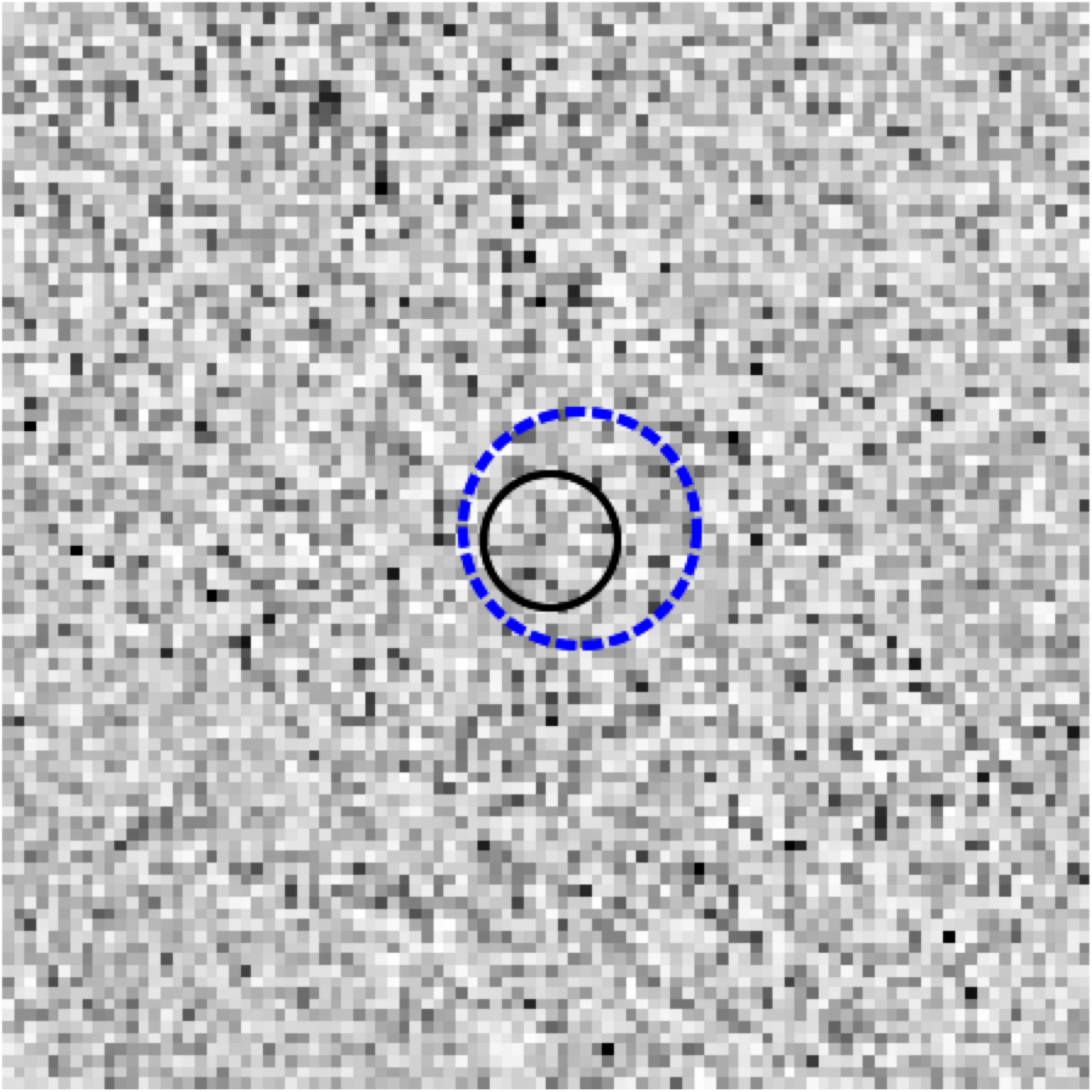}\hspace{-0.1mm}
                \includegraphics[width=2.2cm]{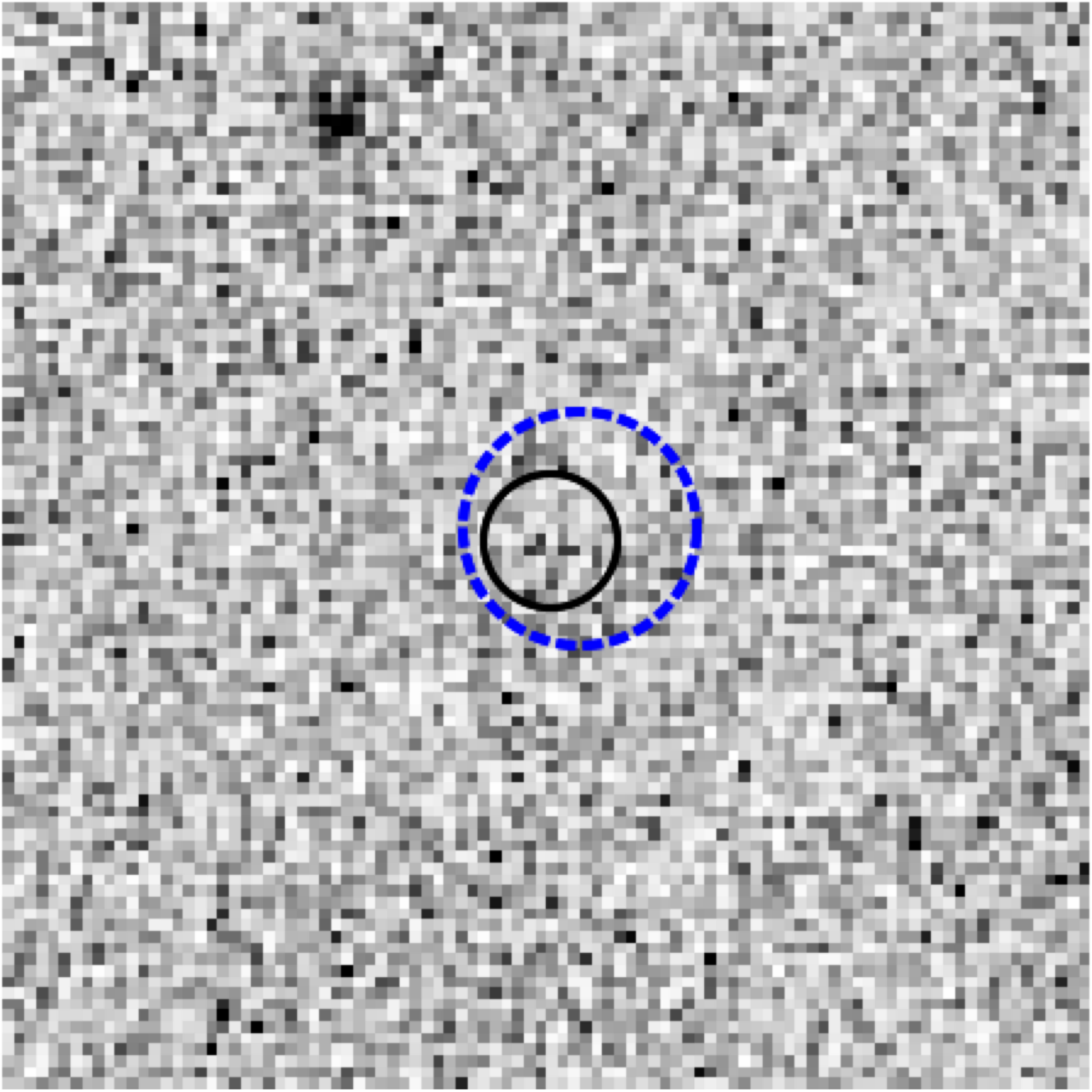}\hspace{-0.1mm}
                \includegraphics[width=2.2cm]{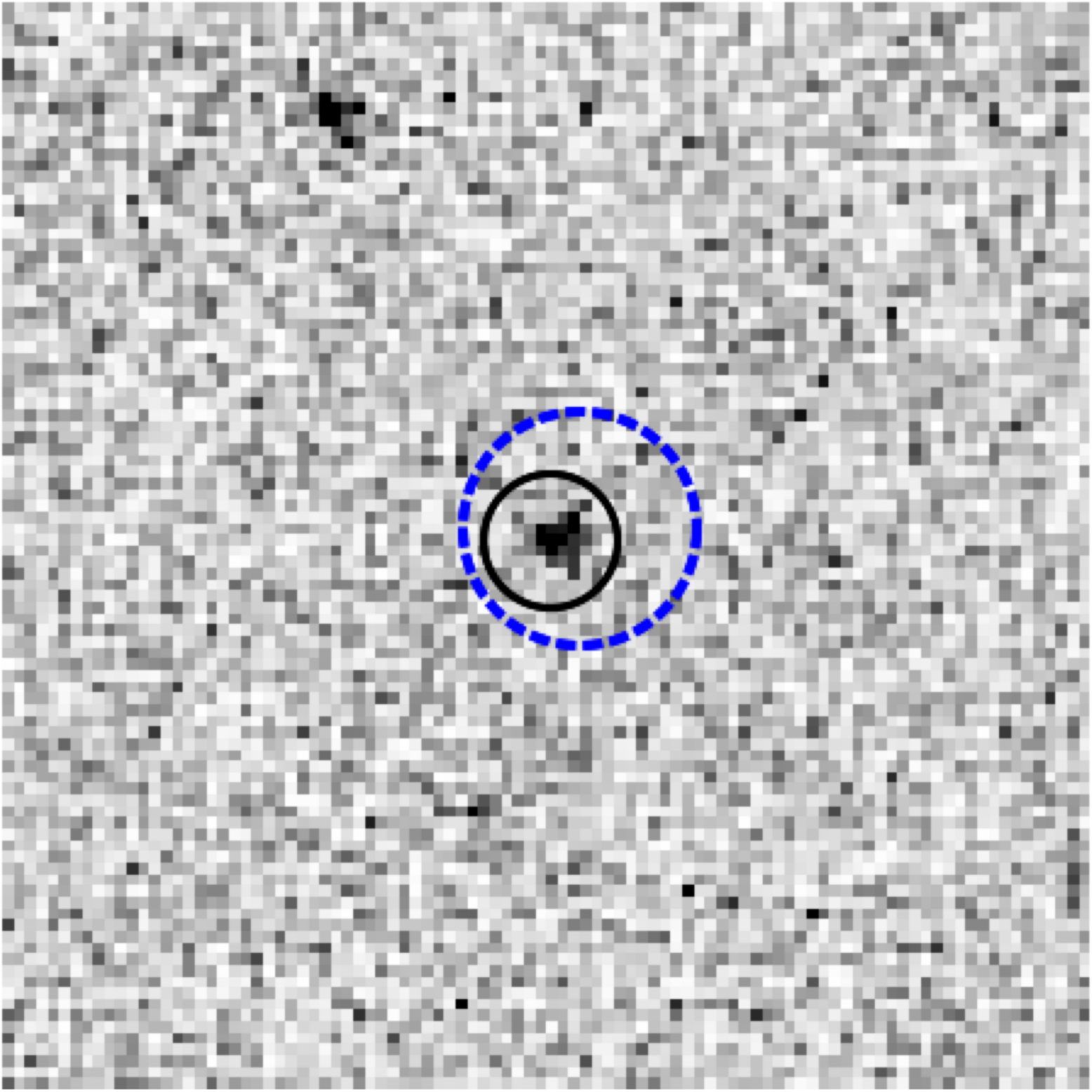}\hspace{-0.1mm}
                \includegraphics[width=2.2cm]{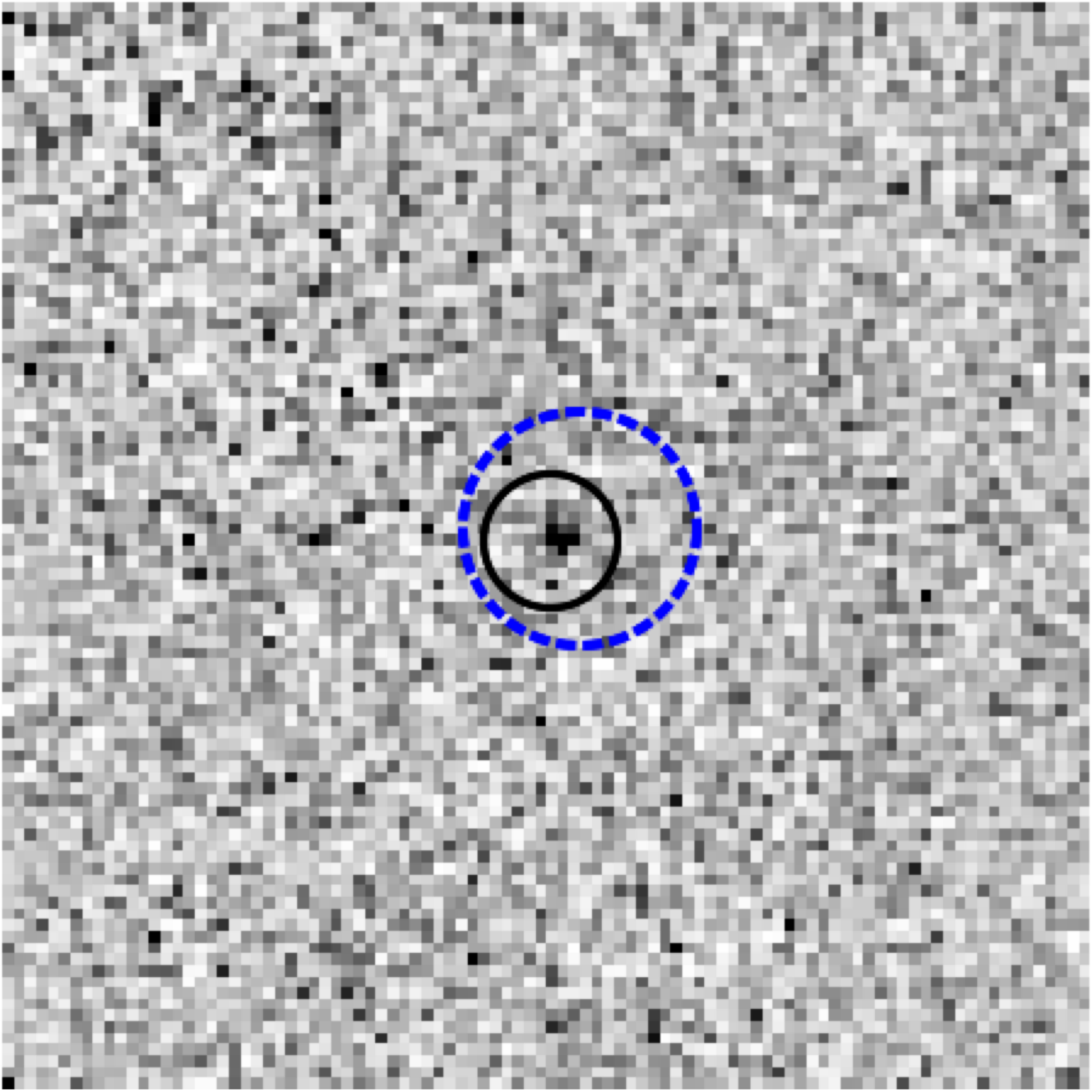}}
        \vskip -0.2 true cm
        \caption{0.4$\arcmin\times0.4\arcmin$ Pan-STARRS PS1 $g,r,i,z,Y$ cutout images of PSO J0309+27. Its optical position is denoted with a black circle that is 1.5$\arcsec$ in diameter. The radio NVSS position is denoted with a blue dashed circle, which is as large as the radio positional error reported in the catalog (2.4$\arcsec$). All images are oriented with north up and east to the left.}
        \label{photimag}
\end{figure*}
Blazars are radio-loud (RL) active galactic nuclei (AGNs) whose relativistic jets are seen at a small angle to the line of sight (Urry \& Padovani 1995).
Since the radiation produced by the jets is strongly boosted and not obscured along the jet direction, blazars are very bright and visible up to high redshifts, ensuring the study of the RL population across cosmic time (see, e.g., Ajello et al. 2009, Caccianiga et al. 2019).
From the space density of blazars it is possible to infer the total space density of RL AGNs, including misaligned and obscured sources that have similar intrinsic physical properties.
Indeed, if we define a blazar as a source observed within a viewing angle equal to 1/$\Gamma$, where $\Gamma$ is the bulk Lorentz factor of the emitting plasma, for each observed blazar we expect to find $\sim$2$\Gamma^{2}$ misaligned RL AGNs whose jets point elsewhere (e.g., Volonteri et al. 2011). 
Therefore the discovery of high-z blazars\footnote{In the following, by the term \textit{blazar} we are referring to quasar-like blazars, that is, flat-spectrum radio quasars (FSRQs) because featureless blazars (BL Lac objects) are not observed at high-z to date (see, e.g., Ajello et al. 2014 and references therein).} ensures the census, free from obscuration effects, of supermassive black holes (SMBHs) in the early Universe and provides strong and critical constraints on the accretion mode, the mass and spin of the first seed black holes (e.g., Kellerman et al. 2016). 
Currently only seven blazars\footnote{The criteria adopted to classify a source as a blazar are not unique and they can be based on both radio and X-ray properties. Therefore we consider all the seven objects classified as blazars in the literature at z$\geq$5.0: J0906+6930 (z=5.47; Romani et al. 2004), J1648+4603 (z=5.38; Caccianiga et al. 2019), J1026+2542 (z=5.25; Sbarrato et al. 2012), J0131-0321 (z=5.18; Ghisellini et al. 2015), J1146+4037 (z=5.005; Ghisellini et al. 2014), J1629+1000 (z=5.00; Caccianiga et al. 2019), and J0141-5427 (z=5.0; Belladitta et al. 2019). Among these, however, only J0906+6930, J1026+2542, and J0131-0321 are confirmed blazars from evidence of Doppler boosting with Very Long Baseline Interferometry observations.} have been observed at 5.0$\leq$z$\leq$5.5, of which the most distant known so far is at z=5.47 (Romani et al. 2004).
At redshift larger than 5.5 only nine RL AGNs have been discovered to date (see Table~\ref{radioq}), but none of these have been classified as blazars.
This is essentially because of the rarity of this type of RL AGNs.
By extrapolating the space density of blazars at 4$\leq$z$\leq$5.5, estimated by Caccianiga et al. (2019), who used the cosmological evolution presented by Mao et al. (2017), we expect to find $\sim$2$\times$10$^{-4}$ blazars every square degree with z>5.5 and mag<21.5.
This means that using the wide field surveys available to date about two to three blazars can be discovered at 5.5$\leq$z$\leq$6.5.
For this reason, we performed a systematic search of z$\geq$5.5 blazars using the widest surveys available up to now; these consist of the NRAO VLA Sky Survey (NVSS; Condon et al. 1998, in the radio), the Panoramic Survey Telescope and Rapid Response System (Pan-STARRS PS1; Chambers et al. 2016, in the optical), and the AllWISE Source Catalog (WISE; Wright et al. 2010; NEOWISE, Mainzer et al. 2011, in the mid-infrared).
By combining these catalogs we selected six blazars candidates at 5.5$\leq$z$\leq$6.5, which we proposed for spectroscopic observations at the Large Binocular Telescope (LBT). 
Among these candidates we found PSO J030947.49+271757.31 (hereafter PSO~J0309+27) the most powerful RL AGN ever discovered at z>6.0. 
It is also the fourth RL AGN observed at this redshift.
In addition, dedicated $Swift$--XRT observations allowed us to find that it is the brightest X-ray AGN detected at z>6. \\
The Letter is structured as follows: in Section \ref{selection} we describe the candidate selection, dedicated optical and X-ray observations, and archival radio data; the results of our work are presented in Section \ref{results}; finally, in Section \ref{conc}, we summarize our results and present our conclusions.
The magnitudes reported in this work are all PSF in the AB system, unless otherwise stated. 
We use a flat $\Lambda$CDM cosmology with H$_0$=70 kms$^{-1}$Mpc$^{-1}$, $\Omega_m$=0.3 and $\Omega_\Lambda$=0.7. 
Spectral indices are given assuming S$_{\nu} \propto$ $\nu^{-\alpha}$ and all errors are reported at 1$\sigma$, unless otherwise specified.
\vspace{-0.3cm}
\section{Candidate selection and observations}
\label{selection}
PSO~J0309+27 is an NVSS radio source selected as a $z$-dropout, typical of z$\geq$5.5 AGNs, in the Pan-STARRS PS1 catalog.
The NVSS is a radio imaging flux limited (S$_{lim}$ = 2.5 mJy) survey
covering $\sim$82\% of the celestial sphere (dec > -40$^{\circ}$), carried out at 1.4 GHz with the Very Large Array telescope with an angular resolution of about 45$\arcsec$ (DnC configuration).
Pan-STARRS PS1 covers $\sim$70\% of the entire sky (dec > -30$^{\circ}$) in five broad photometric bands, $grizY$.
The nominal median co-added catalog depths at a signal-to-noise ratio of 5 are $g$ = 23.3, $r$ = 23.2, $i$ = 23.1, $z$ = 22.3, and $Y$ = 21.4 mag (Chambers et al. 2016).\\
From the entire NVSS catalog, we selected only relatively bright (S$_{1.4GHz}$ $\geq$ 9 mJy) and compact objects to provide an accurate (<10$\arcsec$) radio position and to maximize the selection of blazars; we found 269215 sources. 
We then cross-matched NVSS with Pan-STARRS PS1, using a maximum separation equal to 3\arcsec, to guarantee the reliability of the optical counterparts; we selected only relatively bright optical sources ($z_{PS1}$<21.5) and removed objects on the Galactic plane ($|b|$ $\leq$ 20$^{\circ}$).
The bright radio limit, combined with the PS1 magnitudes, favors the selection of blazars that are characterized by the largest radio loudness\footnote{The radio loudness (R) is the parameter that describes the power of the nonthermal synchrotron jet emission with respect to the accretion disk emission: R = S$_{\rm 5GHz}$ / S$_{ \rm 4400\AA}$ in the source rest frame (Kellermann et al. 1989)} values (e.g., Sbarrato et al. 2013).
The photometric criteria that we used to efficiently select RL AGNs candidates above z=5.5 using the \textit{dropout method} are as follows:
\begin{itemize}
        \item[1)] drop in flux: $i_{PS1}-z_{PS1}$ > 1.0
        \item[2)] blue continuum: $z_{PS1}-y_{PS1}$  < 0.5 
        \item[3)] point-like sources: $z_{PS1}-z_{Kron}$ < 0.05
        \item[4)] $g_{PS1}$>22.5 and $r_{PS1}$>22.5 
        \item[5)] no detection in WISE (W2) or $z_{PS1}-$W2 < 5.0 
\end{itemize}
This last constraint has been placed to minimize the contamination by dust reddened AGNs at z=1-2 (see, e.g., Carnall et al. 2015, Caccianiga et al. 2019). 
After these selection criteria  $\sim$150 objects remained.
Finally we carried out a visual inspection to remove all the possible corrupted frames or frames with no reliable photometry.
From this process, we selected a final sample of six blazar candidates.
By generating catalogs of fake sources located at random radio positions, starting from the original NVSS list, and applying the same filters, we found that the expected number of spurious radio/optical matches is three.
Among the six candidates, three have been observed at LBT: one turned out to be a star (likely a spurious ratio/optical match), one is a low-z galaxy, and the third is PSO~J0309+27.
The spectroscopic follow-up is ongoing for the remaining three candidates.
\vspace{-0.3cm}
\subsection{LBT observation}
\begin{figure}[!h]
        \centering
        \includegraphics[width=9.0cm, height=6.0cm]{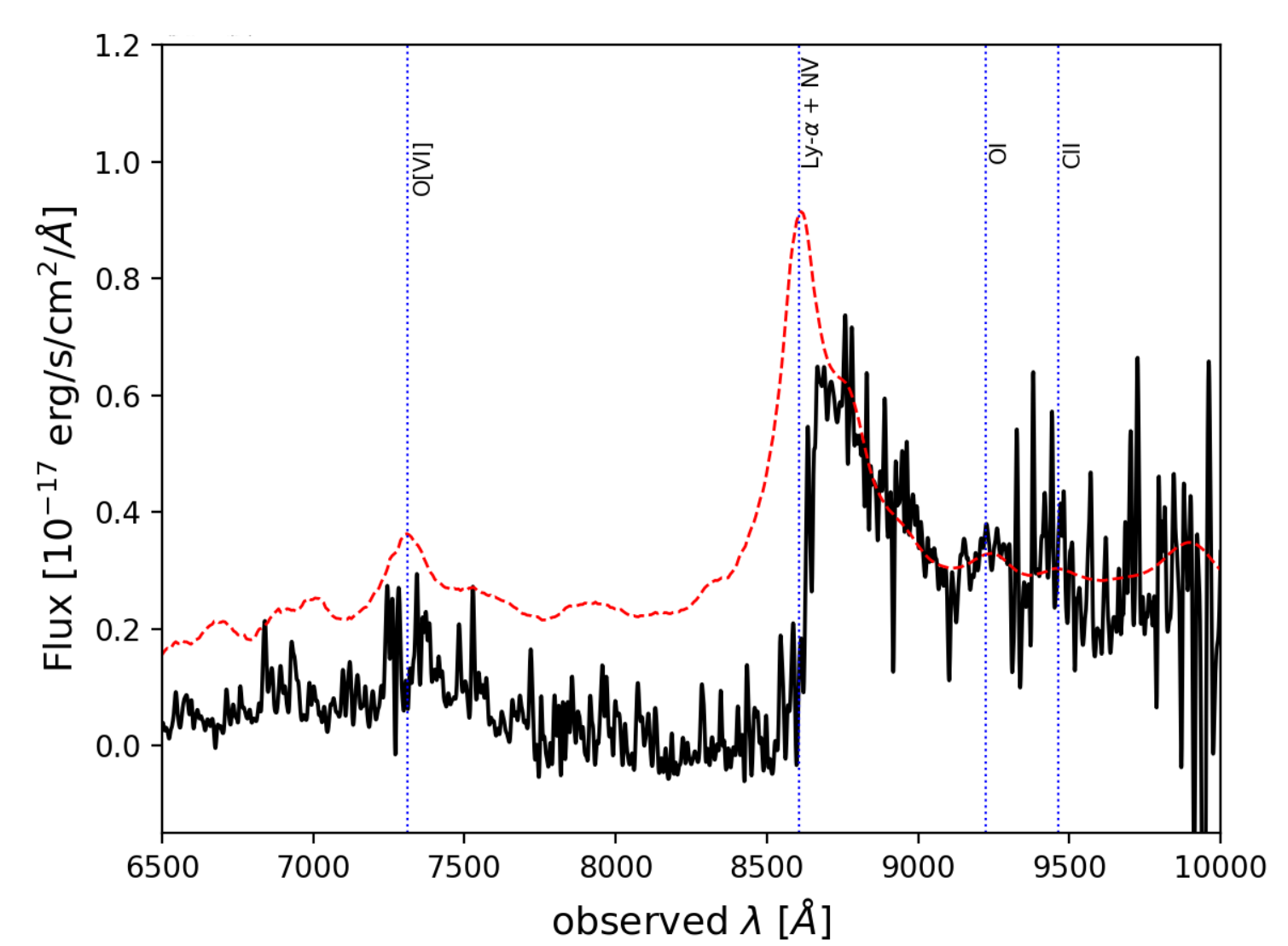}
                        \vskip -0.2 true cm
        \caption{Discovery spectrum of PSO~J0309+27 at z=6.10$\pm$0.03, taken with MODS/LBT. The  O[VI]$\lambda$1033$\mbox{\AA}$, Ly-$\alpha$$\lambda$1216$\mbox{\AA}$,  OI$\lambda$1304$\mbox{\AA}$, and CII$\lambda$1336$\mbox{\AA}$ lines are denoted. The red dashed line represents the quasar template from Vanden Berk et al. (2001) at the redshift of the object for comparison.}
        \label{mods}
\end{figure}

PSO~J0309+27 was selected as a very promising high-z RL AGN candidate from its photometric characteristics, that is, a very low distance between radio and optical position (0.6$\arcsec$) and a high value of $i-z$ (Fig.~\ref{photimag}). 
We confirmed PSO J0309+27 as a high-z AGN at the LBT, the night of the October 2, 2019 (program ID IT-2019B-021, PI S. Belladitta).
We observed it with the Multi-Object Double Spectrograph (MODS; Pogge et al. 2010) in direct mode with the red grating G670L (nominal range: 5000-10000 $\mbox{\AA}$).
We carried out 12 observations of 15 minutes each, with a long-slit of 1.2$\arcsec$ width, for a total observing time of three hours.
The data reduction was performed at the Italian LBT Spectroscopic Reduction Center. 
Each spectral image was independently bias subtracted and flat-field corrected. Sky subtraction was done on 2D extracted, wavelength calibrated spectra.
Wavelength calibration was obtained from spectra of arc lamps reaching a rms of 0.08 $\mbox{\AA}$ on MODS1 and 0.07 $\mbox{\AA}$ on MODS2.\\
The MODS/LBT discovery spectrum of PSO~J0309+27 is shown in Fig.~\ref{mods}.  
Since the Ly-$\alpha$$\lambda$1216$\mbox{\AA}$ line is partially absorbed in the blue side by neutral hydrogen clouds along the line of sight, we estimated the redshift using the  O[VI]$\lambda$1033$\mbox{\AA}$, OI$\lambda$1304$\mbox{\AA,}$ and  CII$\lambda$1336$\mbox{\AA}$ lines (Fig.~\ref{mods}) by fitting them with a single Gaussian profile.
The resulting mean value is z=6.10$\pm$0.03. We verified that the photometric PS1 points are in good agreement with the optical spectrum; the latter has been used to calculate all the optical properties of PSO~J0309+27.
\vspace{-0.3cm}
\subsection{Archival radio data}
\label{archivalradio}
\begin{figure}[!h]
        \centering
        \includegraphics[width=7.0cm]{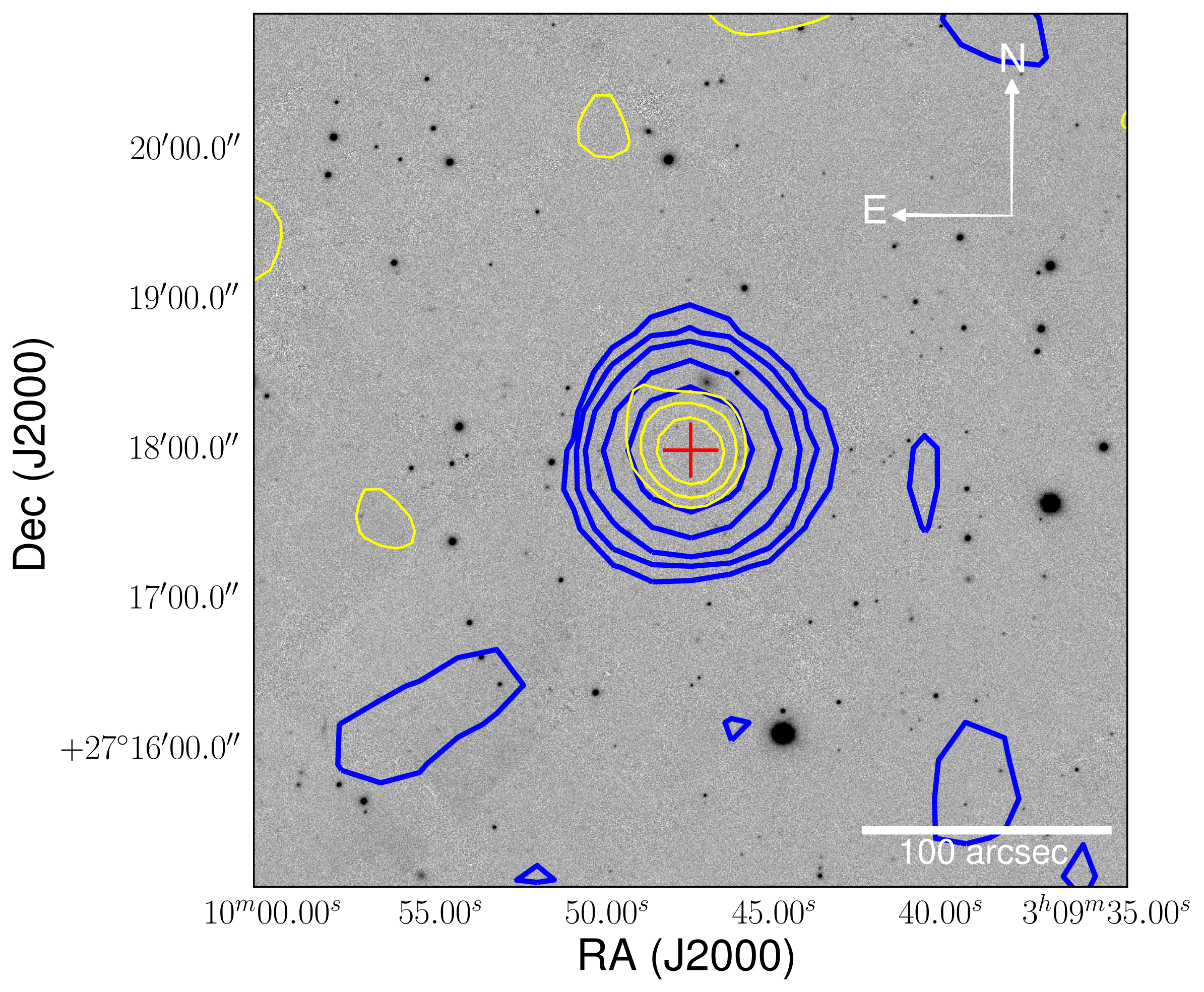}
                        \vskip -0.2 true cm
        \caption{Radio contours from NVSS (blue) and TGSS (yellow) overlapped on Pan-STARRS PS1 $z$–band image (350 arcsec$^2$ field of view). Contours are at (2, 4, 8, 16, 32) $\times$ the off-source rms of each image, which is 350~$\mu$Jy beam$^{-1}$ for the NVSS image and 3.5 mJy beam$^{-1}$ for the TGSS image. The red cross indicates the optical position of the source.}
        \label{contours}
\end{figure}

Aside from the archival NVSS detection at 1.4~GHz, PSO~J0309+27 has also been detected at 147~MHz in the TIFR GMRT Sky Survey (TGSS; Intema et al. 2017).
In Fig.~\ref{contours} we show the overlay of the radio contours on the optical Pan-STARRS PS1 image in $z$-band. 
We found that the radio position at 147 MHz is at $\sim0.5$$\arcsec$ from the Pan-STARRS PS1 coordinates, which is consistent with the 1.4 GHz position.  
From an elliptical Gaussian fit to the radio images, we estimated the integrated flux densities at the two radio frequencies (see Table~\ref{Tmag}), which are consistent with those reported in the two radio catalogs.
Moreover, from the Gaussian fit we found that PSO~J0309+27 is completely unresolved at 1.4~GHz as observed by NVSS, while it is marginally resolved at 147~MHz, showing a slightly elongated morphology (Fig.~\ref{contours}).
This possible extension, which we estimate represents about 10\% of the total flux density, is commonly observed in blazars (e.g., Antonucci \& Ulvestad 1985). 
\vspace{-0.3cm}
\subsection{Swift-XRT observation}
\label{xrays}
\begin{figure}[!h]
        \centering
        \includegraphics[width=5.5cm]{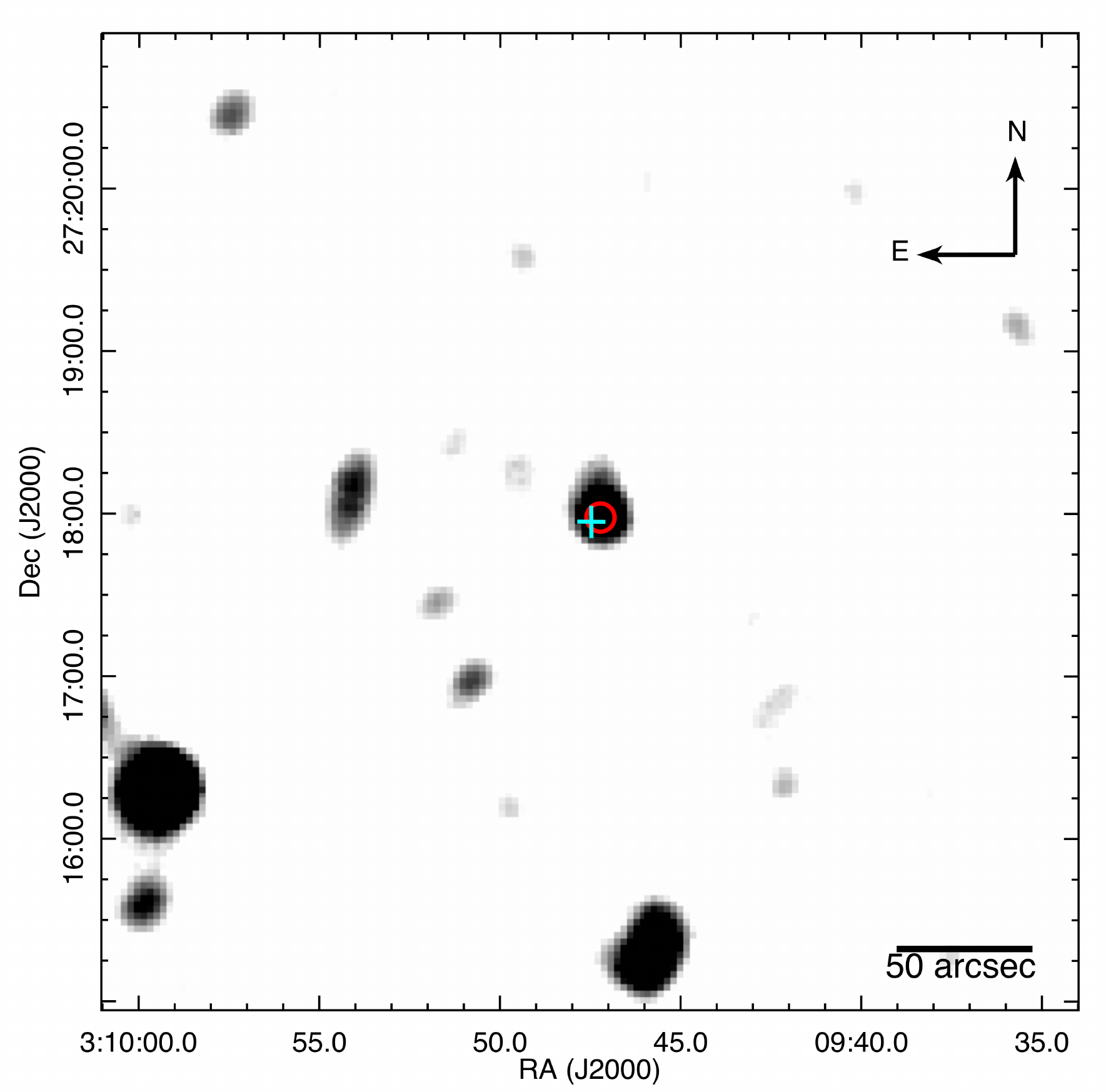}
        \vskip -0.2 true cm
        \caption{ $Swift$-XRT 6$\arcmin$$\times$6$\arcmin$ image of PSO J0309+27. 
        The image is smoothed with a Gaussian filter with $\sigma$=6\arcsec. 
        The X-ray position is indicated with a red circle whose radius (5.2\arcsec) is set equal to the XRT positional error.
        The cyan cross denotes the PS1 position of the source.}
        \label{xray_pos}
\end{figure}
A few days after the discovery we also obtained a $Swift$-XRT (Gehrels et al. 2004) pointed observation that was carried out through two Target of Opportunity (ToO) requests (target ID:12068, P.I.: A. Moretti).
The total exposure time of the observation was 19.1 ks, distributed in eight segments in the period between October 7 and November 15, 2019.
Details about the X-ray analysis are reported in Appendix~\ref{apx}.
The source is clearly detected with a total of 11 counts and with an expected background of 2 counts in the [0.5-10] keV energy band (see Fig.~\ref{xray_pos}).
Running a \textit{wavedetect} algorithm (Freeman et al. 2002) the detection significance is 4.4$\sigma$.
The X-ray position calculated by the \texttt{xrtcentroid} HEASOFT task is 
RA = 03:09:47.22,
dec = +27:17:58.69, with an uncertainty of 5.2\arcsec (90\% confidence).  
This is at 3.8$\arcsec$ from the optical PS1 position.
We fitted data using the C statistics with a simple absorbed power law with the absorption factor fixed to the Galactic value (1.16$\times$10$^{21}$ cm$^{-2}$) as measured by the HI Galaxy map (Kalberla et al. 2005).
Owing to low number of collected photons, the measured spectral photon index is only poorly constrained. We obtained $\Gamma_X$ = 1.6$\pm$0.6, with a corresponding unabsorbed flux of 3.4$^{+5.2}_{-1.9}$$\times$10$^{-14}$ erg s$^{-1}$ cm$^{-2}$ in the observed [0.5-10] keV energy band and a luminosity of 4.4$^{+6.4}_{-3.0}$$\times$10$^{45}$ erg s$^{-1}$ in the [2-10] keV rest frame energy band.
The reported errors on the flux and luminosity take into account both the uncertainty on the photon index and the Poissonian error associated with the photon counts.  
Although the value of the measured photon index is consistent with a flat index that is typical of blazars sources ($\Gamma_X \sim 1.5$; e.g., Giommi et al. 2019), it is also consistent, within 1$\sigma$, with that of radio-quiet (RQ) or misaligned RL AGNs ($\Gamma_X \sim 2;$ e.g., Ishibashi et al. 2010 and references therein).
Therefore, with the current X-ray data, we cannot distinguish between a flat or a steep X-ray spectrum. 
\vspace{-0.3cm}
\section{Multiwavelength properties}
\label{results}
In this section, we compare the radio and X-ray properties of PSO~J0309+27 with those of other AGNs, both RL and RQ, discovered to date at z>5.5.
\vspace{-0.3cm}
\subsection{Radio spectral index and radio loudness}
\begin{figure}[!h]
        \centering
        \includegraphics[width=7.0cm, height=5.5cm]{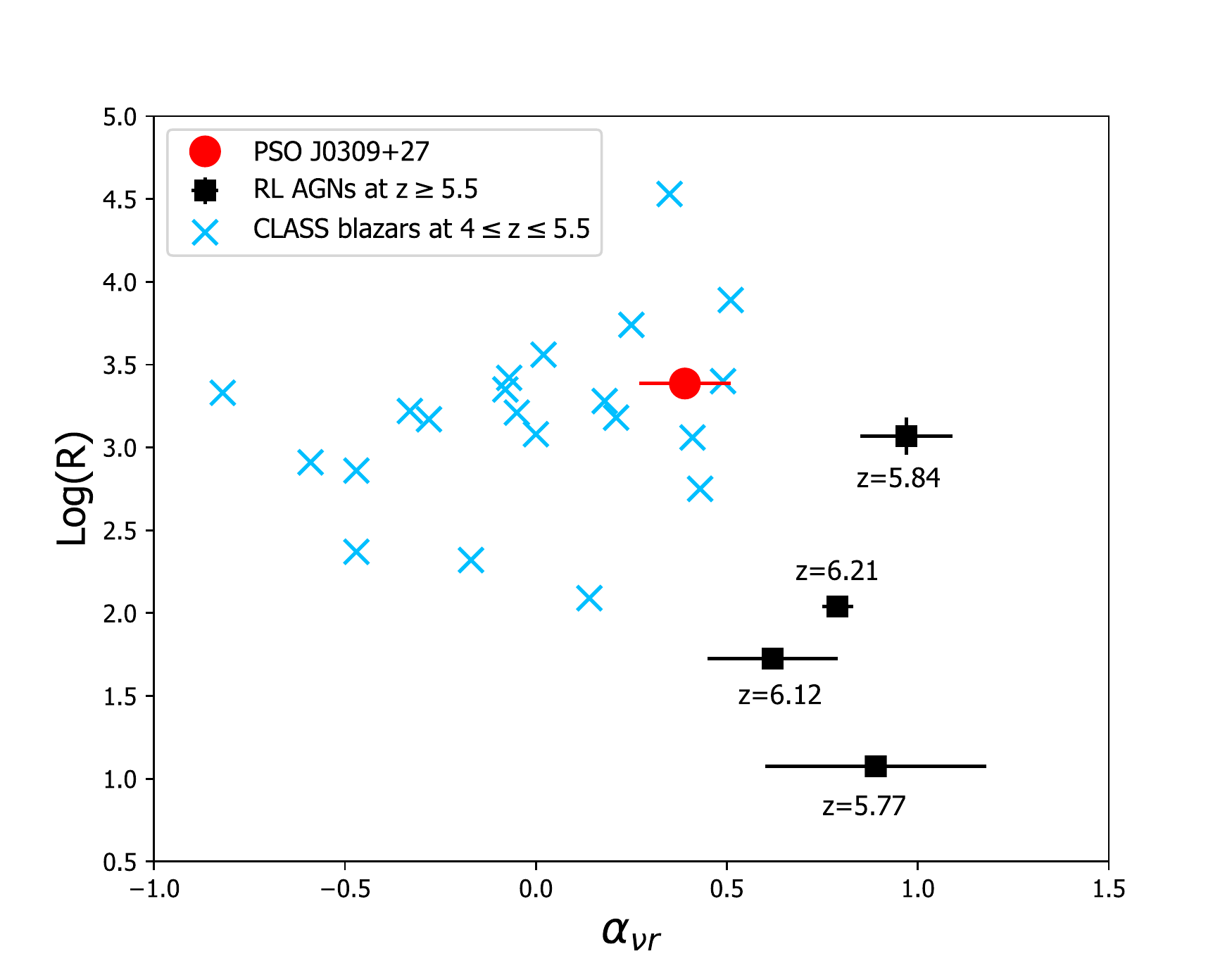}
        \vskip -0.2 true cm
        \caption{Radio-loudness (R) vs. the radio spectral index ($\alpha_{\nu r}$) of PSO~J0309+27 (red point) compared to RL AGNs discovered at z$\geq$5.5 (black squares) for which a measure of $\alpha_{\nu r}$ in the literature is available. For PSO~J0309+27, the value of $\alpha_{\nu r}$ computed from the peak flux densities is shown.
        Light blue crosses represent the CLASS high-z (4$\leq$z$\leq$5.5) blazars.
        Among z$\geq$5.5 RL AGNs PSO~J0309+27 has the highest R ever measured and it is the only one with a flat $\alpha_{\nu r}$.}
        \label{Ralpha}
\end{figure}

From the two radio flux densities (Sect.~\ref{archivalradio} and Table~\ref{Tmag}) and assuming a single power law distribution for the continuum emission (S$_{\nu} \propto$ $\nu^{-\alpha_{\nu r}}$), we computed the radio spectral index of PSO~J0309+27 between 1.4 GHz and 147 MHz, finding $\alpha_{\nu r}$ = 0.44$\pm$0.11.
Using the peak flux densities, which better describe the core emission, we obtained a slightly flatter spectral index ($\alpha_{\nu r}$ = 0.39$\pm$0.12).
This value can be considered flat within the uncertainties ($\alpha_{\nu r}$<0.5; Condon1984) and represents an indication of the presence of a beamed jet  (e.g., Jarvis \& McLure 2006).\\
From the observed optical and radio flux densities we calculated the radio loudness (R) of PSO~J0309+27. 
As mentioned in Sect.~\ref{selection}, a high R supports the idea that the radio emission is boosted along our line of sight. 
We estimated the flux density at 4400 $\AA$ by extrapolating the 1350 $\AA$ flux density from the optical spectrum, assuming a power law continuum with an optical spectral index of 0.44 (Vanden Berk et al. 2001).  
We obtained R=2500$\pm$500, confirming that the radio emission dominates the optical emission.
This high value of R is typical of high-z blazars, as the majority ($\sim$60\%) of blazars found in the literature at z$\geq$4.5 show log(R)>2.5 (Belladitta et al. 2019). 
In Fig.~\ref{Ralpha} we compared the radio spectral index and the radio loudness of PSO~J0309+27 with those of other RL AGNs discovered so far at z$\geq$5.5 for which a measurement of the radio index is available in the literature (Table~\ref{radioq}). 
For comparison we also plotted the values of R and $\alpha_{\nu r}$ of the high-z (4$\leq$z$\leq$5.5) blazar sample of Caccianiga et al. (2019).
We found that at z$\geq$5.5 PSO~J0309+27 has the highest R ever measured and is the only one with a flat radio spectrum. 
Indeed, none of the other RL AGNs at z$\geq$5.5 is classified as blazar, based on high resolution Very Long Baseline Interferometer observations (Frey et al. 2003, 2005, 2008, 2010, 2011; Momjian et al. 2008, 2018; Cao et al. 2014; Coppejans et al. 2016).
\vspace{-0.3cm}
\subsection{X-ray properties}
\begin{figure}[!h]
        \centering
        \includegraphics[width=8.5cm, height=7.0cm]{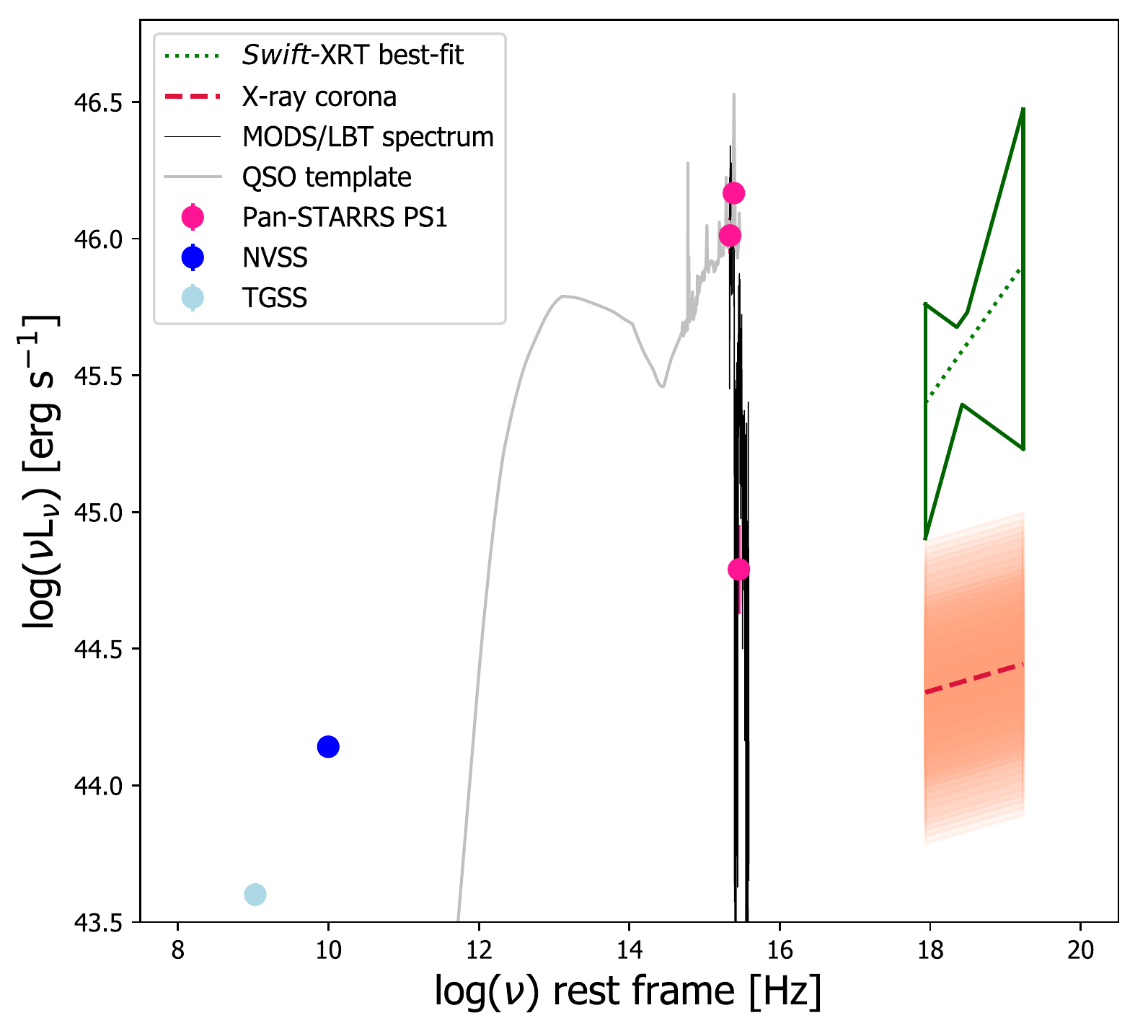}
        \vskip -0.3 cm
        \caption{Rest frame spectral energy distribution of PSO~J0309+27 from radio to X-rays frequencies. In addition to the radio and optical photometric points, in  X-rays the best-fit emission in the observed [0.5-10] keV energy band (green dotted line) with its uncertainty is shown.
        The optical MODS/LBT spectrum (in black) and a quasar template (Polletta et al. 2007; gray solid line) as a guide line are indicated. 
        The red dashed line represents the coronal emission expected from a RQ AGN with the same L$_{2500 \AA}$ of PSO~J0309+27 according to Just et al. (2007) relation; the orange shaded area is the 1$\sigma$ uncertainty on this estimate.}
        \label{sed}
\end{figure}
Blazars, in particular the FSRQ class considered in this work, are characterized by a strong X-ray emission with respect to the optical thermal emission from the accretion disk.
Following Ighina et al. (2019), we computed the  $\tilde{\alpha_{ox}}$\footnote{$\tilde{\alpha_{ox}}$=-0.3026$\log{\frac{L_{10keV}}{L_{2500Å}}}$, both luminosity computed in the source rest frame.} parameter of PSO~J0309+27, which quantifies the relative strength of the X-ray emission with respect to the optical/UV component. 
Unlike the standard $\alpha_{ox}$ parameter (Tananbaum et al. 1979), defined at 2 keV rest frame, the $\tilde{\alpha_{ox}}$ is defined at higher energy (10 keV rest frame), which is more suitable for high-z objects because it is closer to the maximum of the instrumental sensitivity of the most common X-ray telescopes. Therefore, the best-fit normalization is less dependent on the actual spectral shape of the source, which in our case is poorly constrained by the spectral analysis.
We computed the $\tilde{\alpha_{ox}}$ assuming two different values of $\Gamma_X$: 2.0 (typical of z>6 RQ AGNs) and 1.5 (typical of blazars, e.g., Ghisellini et al. 2015).
As expected the value of $\tilde{\alpha_{ox}}$ does not depend significantly on the choice of $\Gamma_X$ and it is equal to 1.09$^{+0.07}_{-0.04}$.
According to Ighina et al. (2019) this value is typical of high-z blazars. 
We also compared the $\tilde{\alpha_{ox}}$ of PSO~J0309+27 with those reported in Vito et al. (2019, hereafter V19), which includes all the z>6 RQ AGNs observed in the X-rays to date.
We want to establish whether the X-ray emission of PSO~J0309+27 is significantly more luminous compared to what is observed in non-blazar AGNs at similar redshifts. 
Indeed in a FSRQ the optical/UV emission has the same origin as in RQ AGNs (the accretion disk), while the X-ray emission is dominated by the boosted jet component that can largely overwhelm the coronal emission. 
To convert the values of $\alpha_{ox}$ reported in V19 into $\tilde{\alpha_{ox}}$ we used the relation of Ighina et al. (2019)\footnote{$\tilde{\alpha_{ox}}$ = 0.789$\alpha_{ox}$ + 0.212($\Gamma_X -1.0$)}. 
We found that PSO~J0309+27 has a much flatter $\tilde{\alpha_{ox}}$ compared to the average value of the V19 sample (1.09 vs <$\tilde{\alpha_{ox}}$> = 1.54), corresponding to an X-ray emission (at 10 keV) a factor $\sim$30 larger, on average, than that of the V19 AGNs, for the same UV luminosity. 
Considering the standard deviation of the values of $\tilde{\alpha_{ox}}$ in the V19 sample ($\sigma$=0.11) and the uncertainty of our measure (0.04), the $\tilde{\alpha_{ox}}$ of PSO~J0309+27 is 3.7$\sigma$ away from the mean value.
Even considering the dependence between $\alpha_{ox}$ and L$_{2500 \AA}$, commonly observed in RQ AGNs, found in Just et al. (2007, see also Strateva et al. 2005, Steffen et al. 2006, Nanni et al. 2017 for similar relations), the value of $\tilde{\alpha_{ox}}$ of PSO~J0309+27 remains significantly ($\geq2\sigma$) flatter compared to z>6 RQ AGNs.
This confirms that PSO~J0309+27 is different from the AGNs currently discovered at z>6 in terms of X-ray emission (see also Fig~\ref{sed}). 
This fact, combined with the very strong radio loudness and the flat radio spectral index, convincingly supports the idea that the radio and X-ray emission of PSO~J0309+27 is produced by a relativistically boosted jet, i.e., that the object is a blazar.
\vspace{-0.3cm}
\section{Discussion and conclusions}
\label{conc}
In this letter we presented the discovery and multiwavelength properties of PSO~J0309+27, the most powerful RL AGN currently known at z>6.0, when the Universe was only $\sim$0.9 billion years old.
This source was discovered by cross-matching the NVSS radio catalog and the first data release of Pan-STARRS.  
The radio, optical, and X-ray photometric and spectroscopic properties described in the previous sections (i.e., flat radio spectrum, very high R, and strong X-ray emission compared to the optical), strongly support the hypothesis that the radio and X-ray emissions of PSO~J0309+27 are jet dominated, i.e., that the object is a blazar (see Table~\ref{Tmag} for a summary of its multiwavelength properties). \\
Assuming that this is the only blazar at this redshift, we can infer the first unbiased (not affected by obscuration effects) measurement of the space density of RL AGNs at z$\sim$6, including the contribution of dust reddened and obscured (Type 2) sources.
PSO~J0309+27 was selected in an area of $\sim$21000 deg$^2$ corresponding to a comoving volume between redshift 5.5 and 6.5 of 359 Gpc$^3$. 
This implies a space density of 5.5$^{+11.2}_{-4.6}$ $\times$10$^{-3}$ Gpc$^{-3}$ in this redshift bin for blazars with an optical absolute magnitude equal or brighter than that of PSO~J0309+27 (M$_{\rm 1450\AA}$ = -25.1).
This value should be considered a lower limit as our spectroscopic follow-up is still ongoing. 
This estimate agrees with the predictions based on the cosmological evolution presented by Mao et al. (2017).\\
Assuming a reasonable value of $\Gamma$ (10, Saikia et al. 2016), we derived a comoving space density of RL AGNs with the same characteristics as PSO~J0309+27 equal to 1.10$^{+2.53}_{-0.91}$ Gpc$^{-3}$.
The total space density of AGNs (both RL and RQ) at the same redshift, obtained by integrating the luminosity function of Matsuoka et al. (2018) over M$_{\rm 1450\AA}$<$-$25.1, is 3.64$^{+0.09}_{-0.03}$ Gpc$^{-3}$.
Therefore RL AGNs may constitute $\sim$30\% of the total AGN population at z$\sim$6. 
Considering the large uncertainties on this estimate, however, the fraction of RL is also consistent with the value of $\sim$10\% typically observed at lower redshifts (e.g., Ivezi{\'c} et al. 2002).\\ 
In order to reduce the statistical uncertainty and establish a possible cosmological evolution of the fraction of RL AGNs with redshift larger samples of blazars at z>6 are needed.
While at the moment the selection of such a sample is not possible, it will be within the reach of the incoming optical and radio surveys performed by the next generation telescopes.
In particular, the combination of the Large Synoptic Survey Telescope (Vera C. Rubin Observatory; Ivezic et al. 2008) data with future radio all-sky surveys (e.g., the Evolutionary Map of the Universe; Norris et al. 2011) will permit the discovery of more than 15--20 blazars at z>6, allowing the most accurate estimate (not affected by obscuration) of the space density of RL AGN population at the end of the re-ionization epoch. 

\begin{acknowledgements}
This work is based on observations made with the Large Binocular Telescope (LBT, program IT-2019B-021). We are grateful to the LBT staff for providing the observations for this object.
LBT is an international collaboration among institutions in the United States of America, Italy, and Germany. 
This work also used data from observations with the Neil Gehrels $Swift$ Observatory (ToO request, target ID:12068).
This work made use of data supplied by the UK Swift Science Data Centre at the
University of Leicester.
We acknowledge financial contribution from the agreement ASI-INAF n. I/037/12/0 and n.2017-14-H.0. 
SB, AM and AC acknowledge support from INAF under PRIN SKA/CTA FORECaST.
CS is grateful for support from the National Research Council of Science and Technology, Korea (EU-16-001).
The Pan-STARRS1 Surveys (PS1) have been made possible through contributions of the institutes listed in https://panstarrs.stsci.edu.
The NVSS data was taken by the NRAO Very Large Array. The National Radio Astronomy Observatory is a facility of the National Science Foundation operated under cooperative agreement by As- sociated Universities, Inc.
GMRT is run by the National Centre for Radio Astrophysics of the Tata Institute of Fundamental Research.
\end{acknowledgements}

\appendix

\section{Tables}
In this Appendix, we report the table containing all the RL AGNs discovered to date at z$\geq$5.5 and the table that describes the multiwavelength properties of PSO~J0309+27. 
Since these properties suggest the blazar nature of the source, in Table~\ref{Tmag} we report the X-ray flux in the observed [0.5-10] keV energy band and the luminosity in the [2-10] keV rest-frame energy band assuming a photon index typical of a blazar ($\Gamma_X$ = 1.5).

\begin{table}[!h]
        \caption{Radio-loud AGNs discovered to date at z$\geq$5.5}
        \label{radioq}
        \tiny
        \centering
        \begin{tabular}{lllll}
                \hline\hline
                name                 & z (Ref) &   S$_{\rm 1.4GHz}$    & $\alpha_{\nu r}$  (Ref) & R                \\
                (1)                  & (2)     &  (3)              & (4)                      & (5)              \\  
                \hline                                                                                                       
                J034141$-$004812       & 5.68 (1) &  2.09            & 0.75$^*$      (1)     & 178.0$\pm$40.5   \\       
                J083643+005453       & 5.77 (2) &  1.11            & 0.89$\pm$0.29   (8)     & 11.9$\pm$0.3     \\
                J090132+161506       & 5.63 (1) &  3.90             & 0.75$^*$       (1)     & 91.4$\pm$8.8     \\
                J142738+331241       & 6.12 (3) &  1.03            & 0.62$\pm$0.17   (8)     & 53.3$\pm$4.1     \\
                J142952+544717       & 6.21 (4) &  2.95            & 0.79$\pm$0.04   (8)     & 109.2$\pm$8.9    \\
                J160937+304147       & 6.14 (1) &  0.48            & 0.75$^*$        (1)     & 28.3$\pm$8.6     \\
                J205321+004706       & 5.92 (5) &  0.43            & 0.75$^*$        (1)     & 44.1$\pm$18.7    \\
                J222843+011031       & 5.95 (6) &  1.32            & 0.8              (1)     & 61.3$\pm$20.9    \\
                J232936$-$152016${^a}$ & 5.84 (7) &  14.90            & 0.97$\pm$0.12   (7)     & 1169.5$\pm$300.5 \\
                \hline
        \end{tabular}
        \vskip -0.2 true cm
        \tablefoot{Col (1): object name; For the object denoted with $a$ the reported values of $\alpha_{\nu r}$ and R are the median of the results of Ba{\~n}ados et al. (2018); Col (2): redshift from the literature and reference; Col (3): observed radio flux density at 1.4 GHz in mJy; Col (4): radio spectral index and reference. For objects denoted with an $*$, $\alpha_{\nu r}$ have been assumed by Ba{\~n}ados et al. (2015); Col (5): radio loudness as defined in Kellerman et al. (1989). List of references: 1=Ba{\~n}ados et al. (2015), 2=Fan et al. (2001), 3=McGreer et al. (2006), 4=Willott et al. (2010), 5=Jiang et al. (2009), 6=Zeimann et al. (2011), 7=Ba{\~n}ados et al. (2018); 8=Coppejans et al. (2017)}  
\end{table}

\begin{table*}[!h]
        \caption{Multiwavelength properties of PSO~J0309+27 (RA: 03:09:47.49 ; dec: +27:17:57.31)}
        \label{Tmag}
        \tiny
        \centering
        \begin{tabular}{p{0.85cm}p{0.5cm}p{1.07cm}p{1.07cm}p{1.07cm}p{0.9cm}p{0.9cm}p{1.0cm}p{0.2cm}p{0.8cm}p{0.7cm}p{0.8cm}p{0.7cm}p{0.6cm}}
                \hline\hline
                redshift      & $i$    & $z$     & $Y$   & S$_{1.4 \rm GHz}$  & S$_{147 \rm MHz}$ & $\alpha_{\nu r}$ & R               & $\Gamma_X$     & F$_X$ & L$_X$ & $\tilde{\alpha_{ox}}$ & L$_{2500\AA}$ &  M$_{\rm 1450\AA}$ \\
                (1)  &  (2)   & (3) & (4)  & (5)  & (6) & (7)   & (8)   & (9)  & (10)   & (11)  &  (12)  & (13) & (14) \\ 
                \hline 
                6.10$\pm$0.03  & >23.1 & 21.37$\pm$0.08 & 20.98$\pm$0.13  & 23.89$\pm$0.87 & 64.2$\pm$6.2 & 0.44$\pm$0.11  &  2500$\pm$500 & 1.5 & 3.6$_{-1.8}^{+2.5}$  & 4.0$_{-1.9}^{+2.8}$ & 1.09$^{+0.07}_{-0.04}$ & 6.1$\pm$0.9  & -25.1   \\
                \hline
        \end{tabular}
        \vskip -0.2 true cm
        \tablefoot{Col (1): redshift; Col (2)-(4): PS1 PSF stacked magnitudes; Col (5)-(6): radio flux densities at 1.4 GHz and 147 MHz in mJy; Col (7): radio spectral index; Col (8): radio loudness; Col (9): fixed photon index; Col (10): X-ray flux at [0.5-10] keV (observed frame) in unit of 10$^{-14}$ erg s$^{-1}$ cm$^{-2}$; Col (11): X-ray luminosity at [2-10] keV (rest frame) in unit of 10$^{45}$ erg s$^{-1}$; Col (12): $\tilde{\alpha_{ox}}$ between 10 keV and 2500$\mbox{\AA}$ rest frame; Col (13): luminosity at 2500$\mbox{\AA}$ in unit of 10$^{30}$ erg s$^{-1}$ cm$^{-2}$ Hz$^{-1}$; Col (14): absolute magnitude at 1450$\mbox{\AA}$.}
\end{table*}

\section{$Swift$--XRT analysis}
\label{apx}
\begin{figure}[!h]
        \centering
        \includegraphics[width=7.5cm, height=5.0cm]{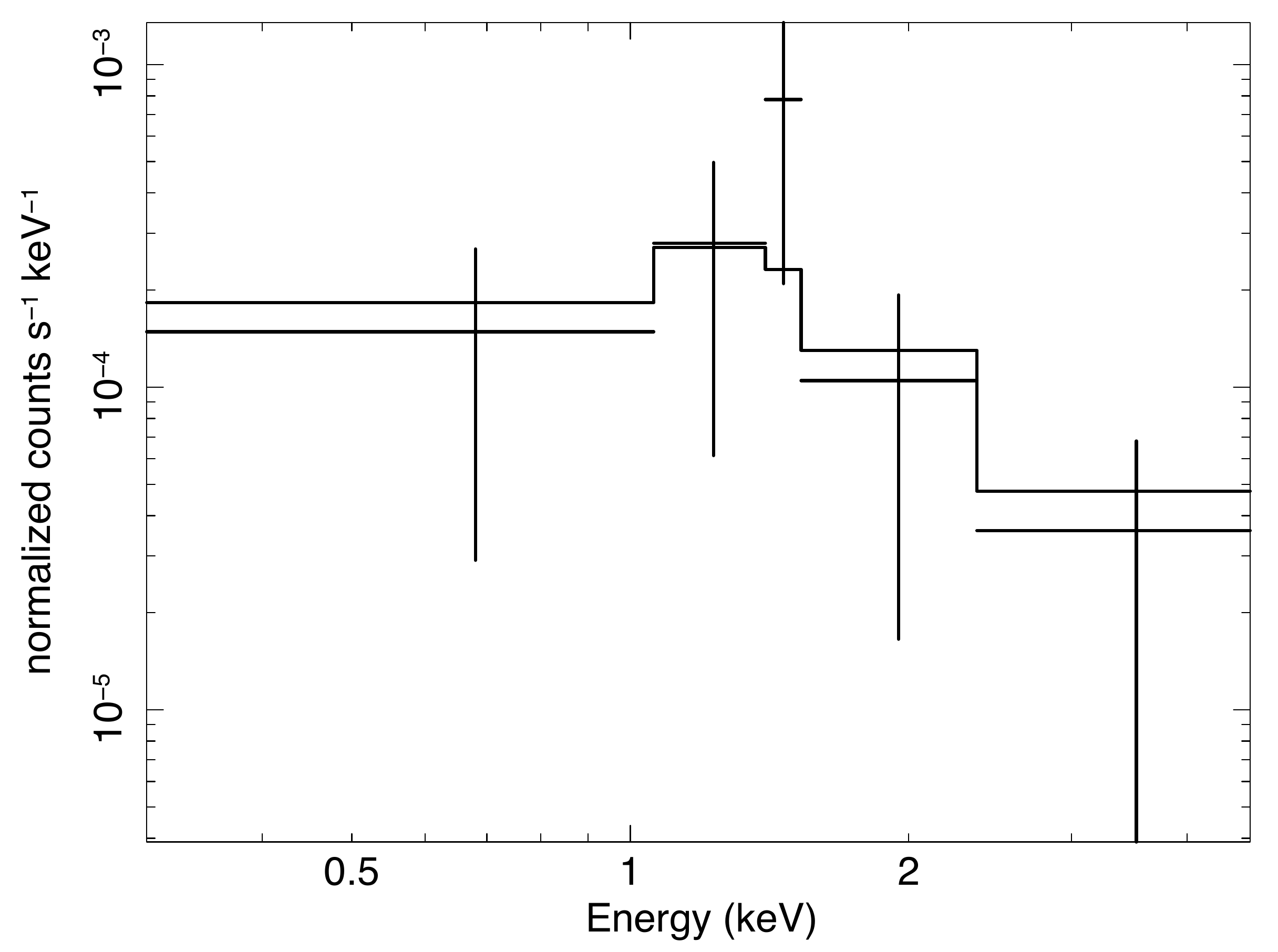}
        \vskip -0.1 cm
        \caption{$Swift$-XRT spectrum of PSO~J0309+27.}
        \label{xspec}
\end{figure}

\begin{figure}[!h]
        \centering
        \includegraphics[width=7.5cm, height=5.0cm]{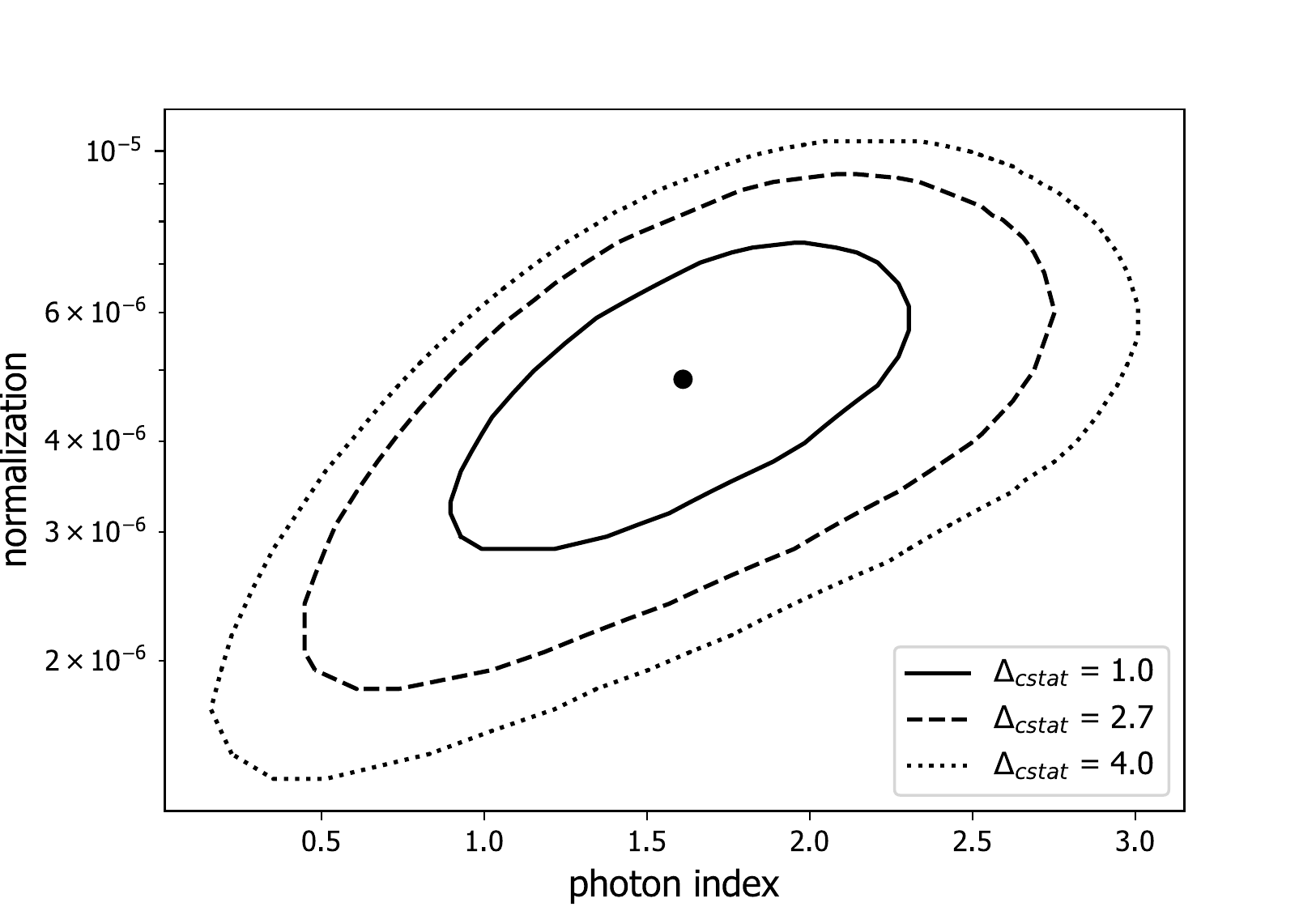}
        \vskip -0.1 cm
        \caption{Confidence contours plot showing the fitted photon index vs. normalization. The contours are drawn for $\Delta_{cstat}$ = 1.0, 2.7, 4.0 respectively, corresponding to 68\%, 90\% and 95\% confidence levels for one parameter. The black point represents the best-fitting value.}
        \label{ellisse}
\end{figure}

In this Appendix, we report all the details on the $Swift$--XRT analysis of PSO~J0309+27. The data were reduced through the standard data analysis pipeline (Evans et al. 2009), running on the UK $Swift$ Science Data Centre web page\footnote{https://www.swift.ac.uk/index.php} using HEASOFT v6.22.
The data for the spectral analysis were extracted from the automatic pipeline, on a circular region of radius of eight pixels, corresponding to 18.8$\arcsec$ ($\sim$75\% of the point spread function). 
Background has been measured in an annulus with an internal radius of 150$\arcsec$ and a external radius of 400$\arcsec$, centered on the source.
A standard spectral data analysis was performed using XSPEC\footnote{https://heasarc.gsfc.nasa.gov/xanadu/xspec/} (v.12.10.1), by fitting the observed spectrum (Fig.~\ref{xspec}) with a simple power law absorbed by the Galactic column density along the line of sight, as measured by Kalberla et al. (2005). Because of the limited number of collected photons, we performed the fit using the C-statistic (Cash 1979) on the data. 
In Table~\ref{xtab} we report the results of the analysis, together with the 68 and 90 per cent confidence ranges for the photon index (see also Fig.~\ref{ellisse}), the observed flux in the [0.5-10] keV energy band, and the luminosity at [2-10] keV rest frame. 
The flux error was computed sampling the parameter pairs (photon index and normalization) within the confidence ellipses and calculating the maximum and minimum of the flux.
We note that the X-ray flux of PSO~J0309+27, at 90\% level of confidence, is larger than 1$\times$10$^{-14}$ erg s$^{-1}$ cm$^{-2}$, confirming its very bright emission, supporting the hypothesis of a boosted emission along our line of sight.


\begin{table*}[!h]
        \caption{Results of the X-ray analysis.}
        \label{xtab}
        \tiny
        \centering
        \begin{tabular}{cccccccc}
                \hline\hline
                source   & $\Gamma_X$  & errors & Flux & errors & Luminosity & errors  & d.o.f. \\   
                (1)  &  (2)   & (3) & (4) & (5) & (6) & (7) & (8) \\ 
                \hline 
                \multirow{2}{*}{PSO J0309+27} & \multirow{2}{*}{1.6} & $\pm$0.6 & \multirow{2}{*}{3.4} & $^{+5.2}_{-1.9}$ & \multirow{2}{*}{4.4} & $^{+6.4}_{-3.0}$ & \multirow{2}{*}{9} \\ 
                &   & $\pm$1.1 & & $^{+9.8}_{-2.4}$ & & $^{+10.8}_{-3.6}$ & \\
                        \hline
        \end{tabular}
        \vskip -0.2 true cm
        \tablefoot{Col (1): source name; Col (2): Best fit value on the photon index; Col (4): X-ray flux at [0.5-10] keV (observed frame) in unit of 10$^{-14}$ erg s$^{-1}$ cm$^{-2}$; Col (6): X-ray luminosity at [2-10] keV (rest frame) in units of 10$^{45}$ erg s$^{-1}$; Col (8): degree of freedom.
        In columns (3), (5), (7) the associated errors at 68\% and 90\% level of confidence are reported.}
\end{table*}
 
\end{document}